\documentclass[journal,onecolumn,12pt]{IEEEtran}
\usepackage{graphicx}
\usepackage{amsmath}
\usepackage{amssymb}
\usepackage{tikz}
\usepackage[utf8]{inputenc}
\usepackage[T1]{fontenc}
\usepackage{lmodern}
\usepackage{float}
\usepackage{dblfloatfix}
\usepackage{cite}
\usepackage{wrapfig}
\usepackage{caption}
\usepackage{subcaption}
\usepackage{lscape}
\usepackage{multirow}
\usepackage{rotating}
\usepackage{setspace}
\usepackage{xcolor}

\ifCLASSINFOpdf
\else
\fi

\begin{document}
\title{Hierarchical Control of Utility-Scale Solar PV Plants for Mitigation of Generation Variability and Ancillary Service Provision}

\author{
Simon~A.~Julien,~
Amirhossein~Sajadi,~
Bri-Mathias~Hodge

\thanks{S.~A.~Julien and B.~M.~Hodge are with the Department of Electrical, Computer \& Energy Engineering and the Renewable and Sustainable Energy Institute at the University of Colorado Boulder, 425 UCB, Boulder, CO 80309, USA, and the National Renewable Energy Laboratory, 15013 Denver W Pkwy, Golden, CO 80401, USA, email: \{Simon.Julien,BriMathias.Hodge\}@colorado.edu, \{Simon.Julien,Bri.Mathias.Hodge\}@nrel.gov}

\thanks{A.~Sajadi is with the Renewable and Sustainable Energy Institute at the University of Colorado Boulder, 4001 Discovery Drive, Boulder, CO 80303, USA, email: Amir.Sajadi@colorado.edu}

}

\markboth{Preprint Draft, \today}%
{~}

\maketitle

\begin{abstract}

This paper presents a hierarchical control system to provide ancillary services from a solar PV power plant to the grid without the need for additional non-solar resources. With coordinated management of each inverter in the system, the control system commands the power plant to proactively curtail a fraction of its instantaneous maximum power potential, which gives the plant enough headroom to ramp up or down power production from the overall power plant, for a service such as regulation reserve, even under changing cloud cover conditions. A case study from a site in Hawaii with one-second resolution solar irradiance data is used to verify the efficacy of the proposed control system. The algorithm is subsequently compared with an alternative control technology from the literature, the grouping control algorithm; the results show that the proposed hierarchical control system is over 10 times more effective in reducing generator mileage to support power fluctuations from solar PV power plants. 
\end{abstract}

\IEEEpeerreviewmaketitle

\section{Introduction}

\IEEEPARstart{E}{lectric} power grids are complex networks in which electrical energy is produced by a diverse set of generation technologies and delivered to energy consumers through transmission and distribution lines. Conventionally, the primary sources of fuel for production of electric power have been coal, oil, natural gas, nuclear, and hydro power. However, in recent years the desire for carbon-free generation has begun a transition towards the integration of 
inverter-based resources (IBR) \cite{8528319}. This transition has accelerated as the cost of energy generation from solar photovoltaic (PV) and wind resources continues to rapidly decline \cite{Lazard2020}. For example, Southwest Power Pool (SPP) recently reported a peak instantaneous renewable generation record of 85.3\% with 97\% of that contribution from wind power. Certainly, this transition is in line with many national energy policies around the world. The Biden administration in the US has set goals to develop a carbon-free power sector by 2035, and a net-zero emission economy by 2050 \cite{Biden}. Similarly, European countries lead by the United Kingdom, Germany, and France at the forefront, alongside many non-European countries such as Japan and  Australia, have established the common goal of releasing net-zero emissions by 2050. Germany has more recently set their goal for net-zero emissions by 2045 \cite{Countries}. While this world-wide trend toward zero-emission power systems is a critical goal in combating the worst effects of climate change, there are many technical challenges that must be overcome to make it a reality.

One of the main challenges with the integration and operation of power systems with increasing levels of  wind and solar PV generation is the variable and uncertain nature of wind speeds and solar irradiance, making them variable renewable energy sources (VRE). Because of this uncertainty in availability, they have not traditionally provided regulation reserves or other forms of ancillary services. Particular to solar PV power plants, the nature of clouds and their rapid formation and movement often creates undesired fluctuations in the plant power output. Because most solar PV power plants have economic incentives to produce the maximum amount of energy, they tend to operate at their maximum possible instantaneous power output, and thus have limited capacity to provide upward reserve products. As a solution, often solar PV power plant developers situate energy storage \cite{olek2014deployment, lee2008small, bravo2018integration} or pair solar PV with fossil fuelled, dispatchable forms of generation \cite{paska2009hybrid, hu2006hybrid, sajadi2019power} to enhance the plant’s flexibility. Combining multiple generation technologies is commonly known as a hybrid power plant and requires significant additional capital costs and complicates market procedures. However, it is expected that in the future solar PV instantaneous penetrations will reach levels at which they will need to be responsible for providing critical grid services. This requires the development of advanced control algorithms for VRE technologies with the advanced functionality required to provide ancillary services without the need for additional resources, which is the focus of this paper. 

A review of the literature suggests that two main directions of research have emerged to address this issue. The first direction focuses on the development of control algorithms that would rely on additional resources bundled with the solar PV power plant, including energy storage and dispatchable generation units to provide flexibility and ancillary services \cite{NELSON2020114963, bohnet2017hybrid, clastres2010ancillary, clastres2010optimal, berrada2016operation, saez2016co, saez2016sizing, saez2016management, lee2018optimal, lee2020sizing, conte2018mixed}. The second direction, which is more recent, focuses on the development of control algorithms for the mitigation of solar PV variability without reliance on additional resources \cite{8370779, vahan, vahan2}. The method developed in \cite{8370779} is an internal forecasting method that would allow a solar PV power plant to forecast cloud coverage and adjust accordingly. This approach is intriguing but comes with the inherent error of probabilistic forecasting and does not provide an algorithm that could be easily constructed into an operational platform. The real-time, grouping control method developed in \cite{vahan} and \cite{vahan2} provides a centralized control algorithm that can react to changes in cloud coverage at second and minute time resolutions. This control technology has recently been integrated into a 141MW Chilean PV power plant \cite{loutan2017demonstration}, which is the first entirely solar PV power plant that is able to bid into ancillary markets without additional non-PV resources. However, the main drawback for this control technology is its central decision-making process that sets a homogeneous setpoint for entire groups of solar PV arrays and neglects that the scattered nature of cloud movements over such large utility-scale PV plants can affect generation at the individual inverter resolution.

This paper advances the state-of-the-art by demonstrating a control algorithm that is able to overcome some of these limitations to provide more accurate power management in solar PV power plants. The control objectives of this control system are to mitigate the power fluctuations caused by the rapid movement of clouds. To this end, it aims to maintain a headroom reserve at each solar array in order to utilize during fluctuation periods. This headroom reserve intrinsically enables the solar power plant to not only follow a generation commitment, but also to respond adequately to real-time dynamic regulation signals from an ancillary service market. The core contribution of this paper is in the development of a hierarchical control system for utility-scale solar PV power plants that enables them to provide ancillary services accurately, even in changing weather conditions. The hierarchical control architecture offers a distributed decision-making process at multiple layers and time scales allowing it to send specific directed control signals to each individual inverter in pursuit of the overall system goal.

To verify the efficacy of the proposed hierarchical control system, a case-study is presented based on data measurements from Oahu, Hawaii. The results from the proposed hierarchical control are then compared with the grouping control method \cite{vahan, vahan2}. The results are presented for two operational functions: (1) following a generation commitment schedule to examine the precision of the algorithm in satisfying the generation level that the plant operator has committed to market and (2) responding to dynamic reserve signals while accurately maintaining and evaluating the status of available headroom for fast reserves. In this paper, PJM's dynamic regulation ancillary service signal (RegD), which includes providing reserves in both the upward and downward directions, from 9am to 5pm, is used as a two second system operator signal \cite{byrne2016estimating}. The results of our simulation show that our hierarchical control algorithm is able to satisfy 99.35\% of the generation commitment and 99.88\% of the ancillary RegD signal requests, whereas by PJM standards and tariffs, a regulation generator responding to a signal needs to comply 75\% of the time \cite{PJM}. Compared to our implementation of the grouping control algorithm \cite{vahan}, our hierarchical approach requires 18 times less mileage when providing regulation reserve and 19 times less supporting regulation reserve when providing a fixed commitment-level output.

\section{Proposed Hierarchical Control System}

\begin{figure}[h]
	\centering
	\includegraphics[width=.6\linewidth]{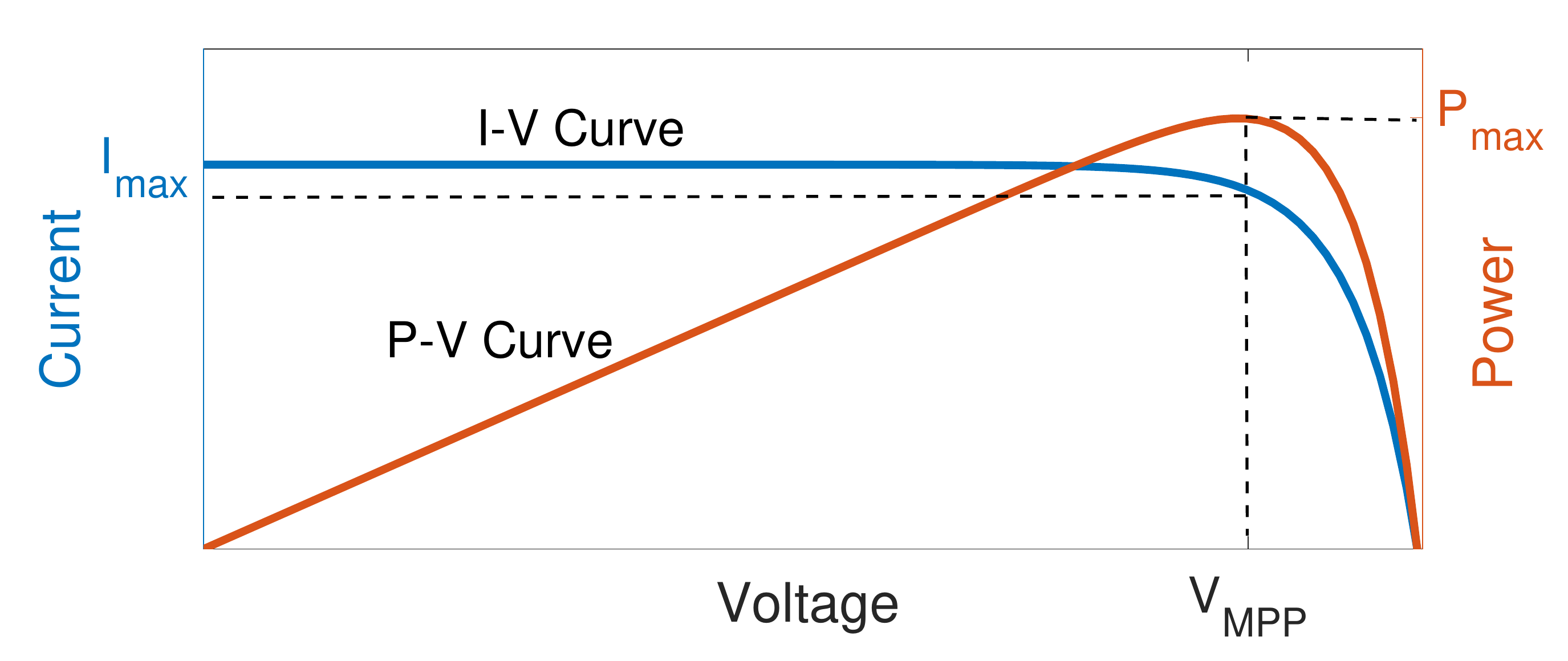}
	\caption{Solar \textit{I}-\textit{V} and \textit{P}-\textit{V} curves showing relationships between voltage, current, and the MPP shown by $P_{max}$}
	\label{fig:IVCurve}
\end{figure}

The core principle behind the proposed control system is the curtailment of power production from the solar PV arrarys. \textit{Curtailing} a variable renewable generation plant means reducing its power output to a specified fraction of its maximum power potential (MPP). Curtailment is achieved by the adjustment of the voltage ($V$) and current  ($I$) levels of solar PV arrays. This is accomplished through a change in the set point of the power electronic inverter which serves as the interface between the solar PV arrays and the power plant collection system, and subsequently, the total power output of a PV power plant will be adjusted \cite{rosu2013practical}, as shown in Fig. \ref{fig:IVCurve}. 

By adjusting the operational set point of the inverter to a fractional value of its maximum potential, enough \textit{headroom} becomes available to ramp up if needed, and regardless of the headroom availability, the production can continue to ramp down from its MPP. This mechanism can be used in a solar PV power plant to enhance operational flexibility and, as the simplest task, the power plant could maintain a flat power output in response to partial shading and rapid cloud movements. Additionally, the available positive headroom can be used to provide ancillary services to the grid by instantaneously adjusting generation power output.

\begin{figure}[h]
	\centering
	\includegraphics[width=.6\linewidth]{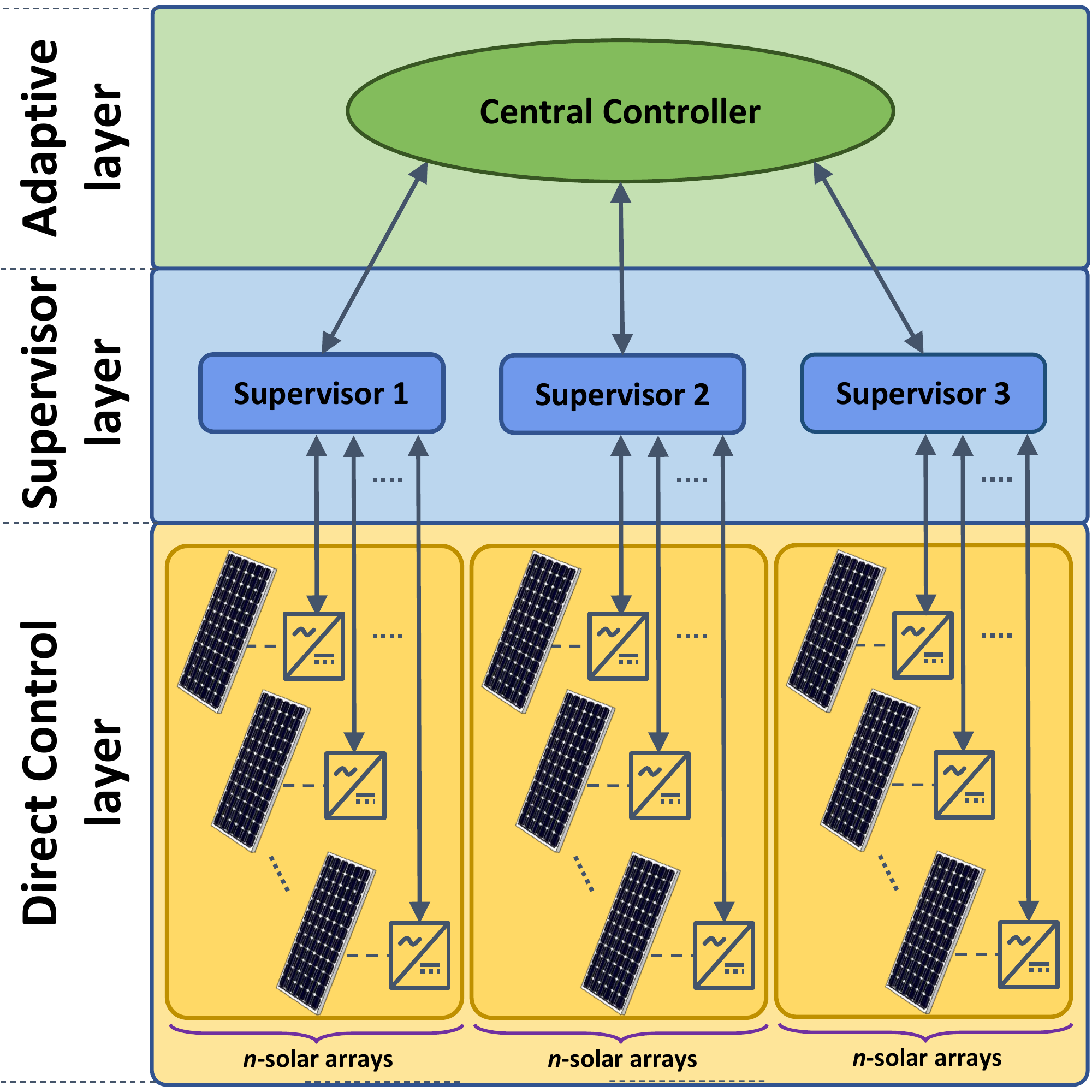}
	\caption{Hierarchical layout of proposed control scheme. Direct controllers on each inverter communicate with their corresponding supervisor controller, and supervisor controllers communicate with the central controller. }
	\label{fig:flow}
\end{figure}

In this paper, three layers of hierarchy for decision-making are considered as depicted in Fig. \ref{fig:flow}. These three layers consists of an \textit{adaptive layer} with a centralized controller at the highest level, a \textit{supervisor layer} with a small number of supervisory controllers as intermediaries, and a \textit{direct control layer} with several inverter controllers at the lowest level of control. Although, only three layers of control are used in this paper, one large benefit of hierarchical control is the modular nature that would allow for adding more supervisor layers (e.g. a super-supervisor layer) in order to effectively expand and coordinate control to more than one power plant. The following provides the decision-making processes, individual responsibilities, and control goals of each control layer.

\subsubsection*{Direct Control Layer}
This layer is the lowest level of control consisting of several control agents that directly regulate voltage and power output of each solar PV inverter at the device level. This layer initially computes the estimated inverter maximum power potential (IMPP) of an inverter and then determines if the IMPP is above the requested power output. If the inverter cannot produce the requested power output, it will ask for ``help" from its supervisor controller. Using typical signaling hardware these computations should be able to send out updated signals in less than 1 second.

\subsubsection*{Supervisor Layer}
This layer works as the ``middle-man" of control. Supervisors first survey the performance of their ``control agents" (corresponding to direct controllers). If any one of the control agents has been flagged for needing ``help", then the supervisor attempts to solve the missing amount of power generation by finding another one of it's control agents to provide additional power production. If all the constituent control agents of a supervisor are operating at IMPP and the cluster is still experiencing a power production deficit, then the supervisor signals to the above layer (adaptive central controller) and requests support from other supervisory clusters. Once the entire system has found a feasible solution, the supervisor signals the final curtailment set points down to each of its control agents. Signal requests could operate within a 10 second window with typical signaling equipment.

\subsubsection*{Adaptive Layer}
This layer uses one centralized controller to receive the external ancillary services signal for power generation, and command the lower tier controllers to find a global solution for power production while considering partial shading impacts. This upper tier communication requests additional production from supervisor clusters that have enough available potential to help compensate for shaded supervisor clusters. Final curtailment set points are then communicated from the adaptive layer to the supervisory layer, to the direct control agents, respectively. As the final stage of this iterative signaling, controllers at this level will be operating at less than 1 minute resolutions.

\begin{figure}[h]
	\centering
	\includegraphics[width=.7\linewidth]{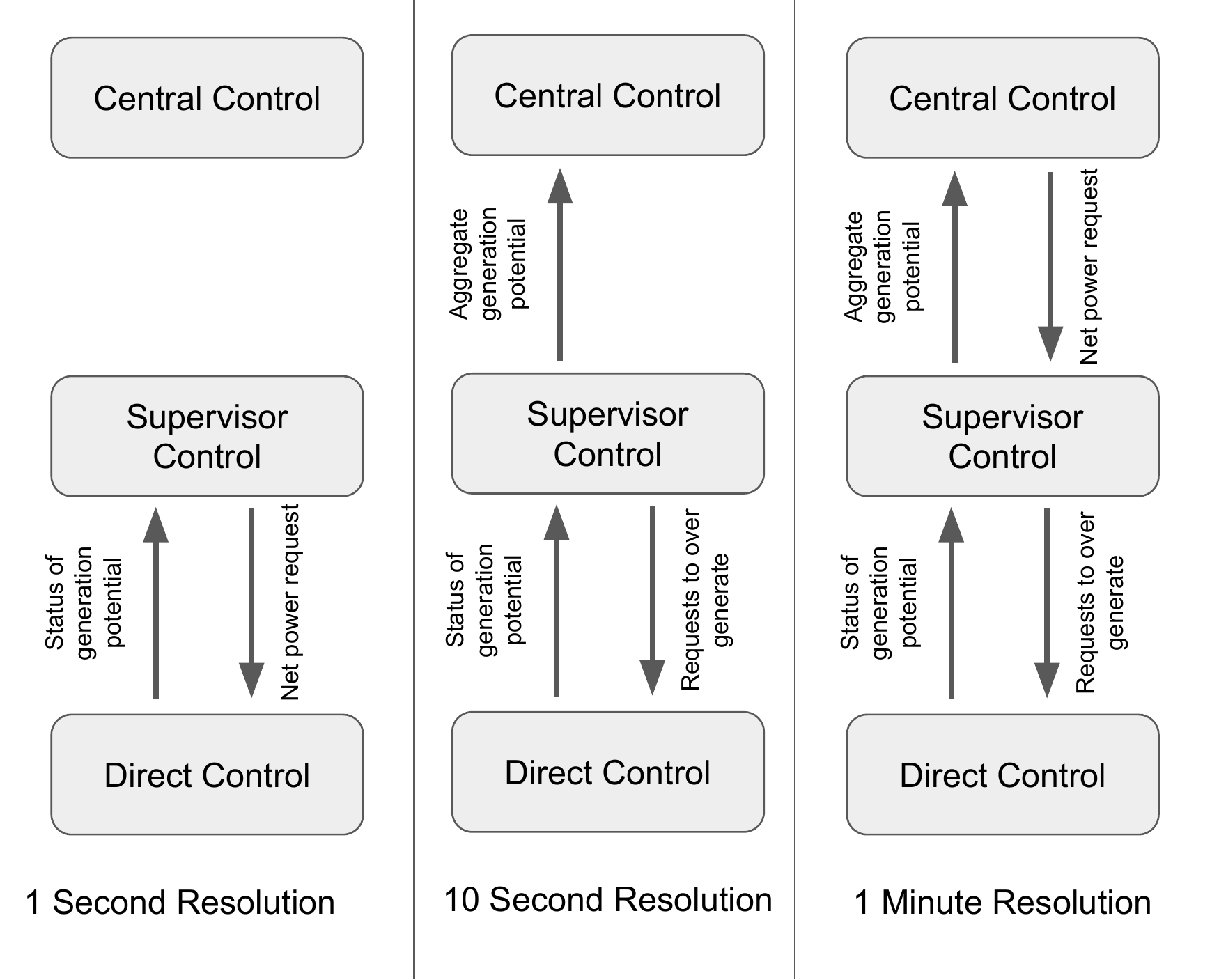}
	\caption{Exchanges of requests between the hierarchical layers and the upper-bounds of their computational decision-making timescales.}
	\label{fig:Layers}
\end{figure}

{
	A great advantage of the hierarchical control architecture is its ability to coordinate the PV system operation by more than one decision-making center. Necessary information is available to each controller at different time frames and at different rates, which allows for less computational intensity while providing the opportunity for the expansion of communications between multiple solar power plants. Dispersing control into a hierarchy of micro-controllers also increases the resiliency of the system, and it can even be implemented with simple, off-the-shelf computing systems. Fig. \ref{fig:Layers} breaks down the communication and the exchange of requests between the hierarchical layers, while also depicting the chronological order of all the information transfer and the corresponding decision-making timescale.}

After the hierarchy of control is established and installed, the layers exchange information formulaically to compute the optimal set point for curtailment at each inverter.
The proposed control algorithm for information exchange and computation involves the following iterative steps: 

\begin{table}[h]
	\begin{tabular}{ll} 
		Step 1: & Estimate System IMPP \\
		Step 2: & Compute Initial Power Output for Controllers \\
		Step 3: & Conduct Correlative Neighbouring \\
		Step 4: & Solve \& Reassign Curtailment Levels \\
		Step 5: & Return to Step 1
	\end{tabular}
\end{table}

\subsection{\textit{Step 1}: Estimate System IMPP}
In the first step the IMPP is computed for each inverter iteratively at computational speeds much faster than changes in solar availability, which is usually over a minute. Hence, there is very little variation in IMPP between iterations, which greatly reduces the error in IMPP computation that is managed at the direct control layer.

\subsubsection*{Formulation}
To evaluate the IMPP of the $i$-th inverter at $t$-th iteration, represented by a subscript index of $i$ and superscript index of $t$, the algorithm must take in the desired fractional curtailment level, denoted by $\alpha_{i}^{t-1}$, and final (after curtailment) inverter power output, denoted by $P_{final,i}^{t-1}$, of the previous iteration, represented by superscript index of $t-1$, as arguments. Thus, the estimation algorithm for the IMPP, represented by $P_{IMPP,i}^t$, at iteration $t$ can be described by:
\begin{equation}
\begin{split}
\label{eq:InvIMPP}
P_{IMPP,i}^t&=P_{IMPP,i}^{t-1}\\
&=\frac{P_{final,i}^{t-1}}{\alpha_{i}^{t-1}}.
\end{split}
\end{equation}

Next, the total plant MPP, $ P_{MPP,system}^{t}$, is computed by simply summing across all $N$ inverters by:
\begin{equation}
\label{eq:SysIMPP}
P_{MPP,system}^{t}=\sum\limits_{i=1}^N P_{IMPP,i}^t
\end{equation}

\subsection{\textit{Step 2}: Compute Initial Power Output for Controllers}
\textit{Headroom} is a positive value that indicates the difference between IMPP and the curtailed operational level of a solar inverter. After the IMPP estimation, the same amount of power generation is requested of all inverters, defined by power request point and denoted by $P_{request}$. This request point is the difference between the expected generation level and the average IMPP of all inverters in the system if all inverters were exposed to the same amount of solar irradiance. The power set point then is compared with the actual available IMPP at each individual inverter and each supervisor controller. The difference between the power set point and IMPP constitutes the residual generation signal. If the residual generation signal for an inverter returns a negative value, the requested generation amount ($P_{request}$) is larger than the IMPP, and the controller will be flagged as ``in need of assistance." 

\subsubsection*{Formulation}
The initial uniform power request to all inverters (equivalent to if all inverters were exposed to the same exact amount of solar irradiance), denoted by $P_{request}$, is computed using the arguments of the desired system power output, represented by $P_{desired}$, and the number of inverters ($N$) as:
\begin{equation}
\label{eq:req}
P_{request}=\frac{P_{desired}}{N}
\end{equation}

The residual power at the $i$th inverter, represented by $P_{res}$, is the difference between the inverter IMPP value, $P_{IMPP,i}$, from the expression \eqref{eq:InvIMPP} and the initial power request, ($P_{request}$, in \eqref{eq:req}). Next, these values are used to compute the index for the need for assistance for each inverter by simply subtracting the initial power request from the inverter IMPP value by:
\begin{equation}
\label{eq:invHeadroom}
P_{res}=P_{IMPP,i}-P_{request}
\end{equation}
where the sign of the value of $P_{res}$ can be used as a reference to determine the state of power generation by the $i$th inverter and can be interpreted using the simple following rule: 
\begin{equation*}
\begin{split}
P_{res}\geq0 \xrightarrow{}& \text{No need for assistance}\\
P_{res}<0 \xrightarrow{}& \text{In need for assistance}
\end{split}
\end{equation*}

The magnitude and sign of $P_{res}$ is sent to the supervisor controllers. The aggregated power output operated by each supervisor controller, denoted by $P_{sup}$, with $N_{agents}$ constituent inverters, is computed similarly by summing $P_{res}$ values across $N_{agents}$ ``agent" inverters. With more than 3 control tiers, the same procedure would be repeated for all higher tiers of control up until the centralized controller is reached.

\subsection{\textit{Step 3}: Conduct Correlative Neighboring}
Thus far, the need for help by individual direct control agents and supervisors has been established. The next natural step is to develop an augmented mechanism to systematically determine the signalling for the assistance procedure. Here a signaling mechanism is proposed with a goal to maximize the probability of receiving support from the inverters initially asked for additional support. To this end, a correlation matrix is trained by historic data in the location to describe the hourly correlation of headroom availability among all inverters and per row. 

Correlations behind inverter behaviors were segmented by the hour as seen in Fig. \ref{fig:CorrMat}, however the resolution of updating these correlation matrices may fluctuate at different geographic locations. The diagonal of the correlation matrix will always return 100\% correlation because an inverter will have perfect correlation with itself. Upon the need to request help to reach its power set point, the controller will go through the list by reaching out first to the least correlated inverters, i.e. those inverters which have the highest probability of not experiencing similar power deficits. Such arrangement of all the controllers within a hierarchical algorithm will greatly affect how ``high up the chain" inverters will have to ask for help. For example, geographically neighboring inverters will likely have very similar solar fluctuation patterns because a passing cloud may cover both of them simultaneously. The hierarchical controller will choose virtual neighbors of controllers that have uncorrelated -- or (ideally) inversely correlated -- solar fluctuations. 

The term \textit{virtual neighboring} is used to suggest that the controllers are neighbors through the eyes of the signaling hierarchy of the controller, but they are not physically neighboring. This will make the support process much more efficient and less computationally expensive as most controllers likely will find an equilibrium solution at the supervisory layer without consulting the adaptive layer.

\subsubsection*{Formulation}
In this study, it was found that the best metric to determine statistical correlation between all inverters is to perform a statistical analysis using irradiance data with granularity of 1 second resolution. This analysis established a $n$ by $n$ correlation matrix for the set of $n$ inverters and each inverter will have a per-unit rating of their level of behavioral correlation to all the other inverters in the plant. Each correlation coefficient $C$ is calculated by the Pearson's Linear Correlation Coefficient where two time-series vectors $X$ and $Y$ are inputted to the function: 

\begin{equation}
\label{eq:Corr}
C(X,Y)=\frac{\sum\limits_{i=1}^t (X_i-\overline{X})(Y_i-\overline{Y})}{\sqrt{\sum\limits_{i=1}^t (X_i-\overline{X})^2\sum\limits_{i=1}^t(Y_i-\overline{Y})^2}}
\end{equation}

The smaller the value of correlation, C(X,Y), the more suitable the neighboring of inverters $X$ and $Y$. Using the final statistical characterization of the solar variation between power plant inverters from the correlation matrix, hierarchical tiers with non-similar or inversely correlated fluctuation patterns can be chosen as virtual neighbors for an improved response.

\subsection{\textit{Step 4}: Solve \& Reassign Curtailment Levels}
By this stage, the available power of each controller has been determined as an indicator of their need for support from the other agents. A mechanism also has been established to determine the most efficient inverter or cluster to reach out to for support. Starting at the supervisory level, any supervisory controller(s) that needs help then notifies the central controller at the adaptive layer and informs them how much power support they need. Subsequently, other supervisor controllers with positive headroom values then ramp up their power production to provide support. Once the power output of each supervisory cluster has been assigned, the process is recursively called to assign individual power curtailment operation levels for every inverter under each supervisor. Once this procedure is complete, the algorithm can settle on a final set point for each inverter that will accurately aggregate to the desired power plant output.

\subsubsection*{Formulation}
The amount of power required to ramp up to provide support, denoted by $P_{help}$, is the cumulative sum of all power generation error signals such that $P_{res}<0$. The algorithm then adjusts the desired power production level of each controller as long as it does not exceed the IMPP. The control logic used here is recursive at each level of control throughout the hierarchical control algorithm to update the operational level of the controller. The
threshold for the iterative process to be considered converged is when $P_{res}\geq0$ for all direct controllers and thus a stable equilibrium is reached. If the power plant has a total aggregated IMPP value above the desired final power output, then the condition $P_{res}\geq0$ will be achievable for all inverters $i$. In the final state of this system, it should be true that the desired system power output is exactly equal to the sum of the final operation of each individual inverter, $P_{final,i}$.

The final step of the iteration is to solve for the curtailed operational efficiencies ($\alpha_i$) of each inverter $i$ as:

\begin{equation}
\label{eq:eff}
\alpha_i=\frac{P_{final,i}}{P_{IMPP,i}}
\end{equation}
The $\alpha_i$ then is used to adjust the voltage ratio at the $i$th inverter. Next is to begin the subsequent iteration, back to Step 1.

With the hierarchy of control established, and the formulaic communication and signaling of the proposed algorithm specified, the next sections demonstrate how the designed hierarchical control algorithm performs in a simulated case study.

\section{Case Study Oahu, Hawaiian Airport}


\subsection{Solar Irradiance Data}
\begin{figure}[h]
	\includegraphics[width=0.95\linewidth]{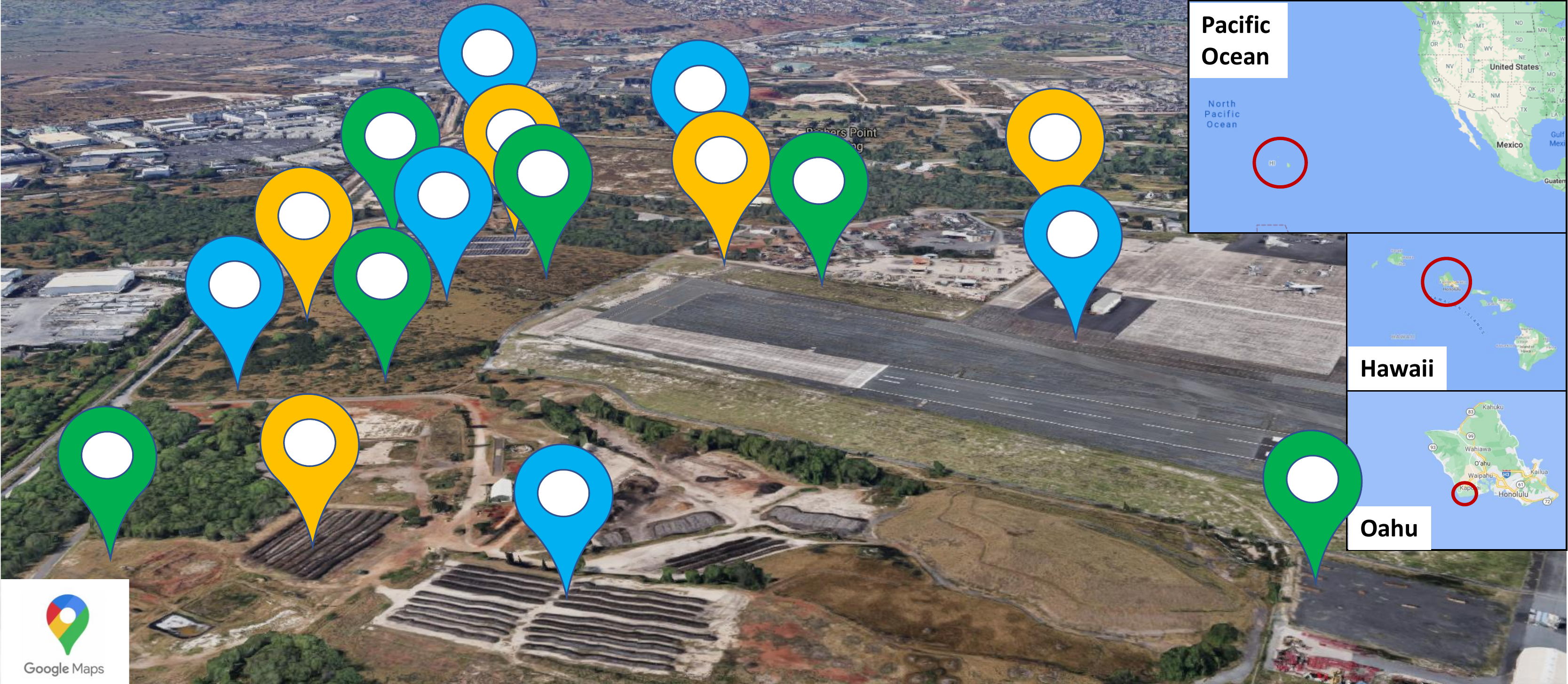}
	\centering
	\caption{\label{fig:map} Satellite image of the distribution of 17 solar irradiance measurement devices around an airport in Oahu, Hawaii \cite{data}. Virtual neighboring clusters are indicated by colors: Blue pins are cluster 1, Green pins are cluster 2, and Orange pins are cluster 3.}
\end{figure}

The measurement data from solar irradiance in Oahu, Hawaii, US, are used as a case study in this paper. This data has 1-second resolution global horizontal irradiance measurements from 17 devices around an airport on this island. These irradiance meters measured starting at 5am and ending at 8pm, every day, April through October, in both 2010 and 2011. In Fig. \ref{fig:map}, the layout of the measured data is presented visually through a satellite image with solar PV graphics marking the location of each irradiance meter. 

\subsection{Solar Irradiance Data to AC Power Data Conversion}

The raw data used in this case study was solar irradiance measurements in $W/m^2$. For the purpose of this study, this data was converted into electric power output of hypothetical solar arrays. Intuitively, the physical power output from any solar array is dependent on the amount of solar irradiance to which the panels are exposed. To convert the irradiance data into a time series of potential AC power output equivalent solar arrays, the Sandia National Laboratory's PVLib package \cite{PVLib2} was used. This class also allows for input of temperature degradation and wind speeds, but these values were left in their default state. given that they are secondary impacts on power output. This conversion approximates an equivalent of a 8MW PV solar power plant with arrays spread across 1.5 square miles.

\subsection{Implementation}

The solar PV power plant and the proposed hierarchical controller were implemented in MATLAB Simulink R2020b. The overall system has 17 direct controllers, 3 supervisory controllers, and 1 central controller. To operate all 17 inverters, the controller setup included two clusters associated with 6 inverters (green and blue in Fig. \ref{fig:map}) and one cluster associated with 5 inverters (yellow in Fig. \ref{fig:map}). These clusters have been selected such that the correlation between inverters within a cluster is at a minimum, so the probability of available headroom within a cluster is maximized.

The controller is implemented to follow the hierarchical process as described in the previous section. Communication signals are exchanged between controllers and layers (with communication delays neglected because of the significant time separation), and within each control MATLAB scripts are written to make the computational measurements and decision making necessary to interpret the status of their generation abilities. Once all inverters in the system have been assigned their final operational level, that value is recorded, and the next iteration begin immediately. With a 2GHz quad-core processor, this simulation took 14.144 seconds to simulate 54000 seconds of operation. Assuming simulation time will be much faster than real-time processing and communication, a breakdown of the maximum amount of processing time we anticipate using common computational hardware is shown in the column labels of Fig. \ref{fig:Layers}.

\section{Results and Discussion}

\begin{figure}[h]
	\centering
	\includegraphics[width=.7\linewidth]{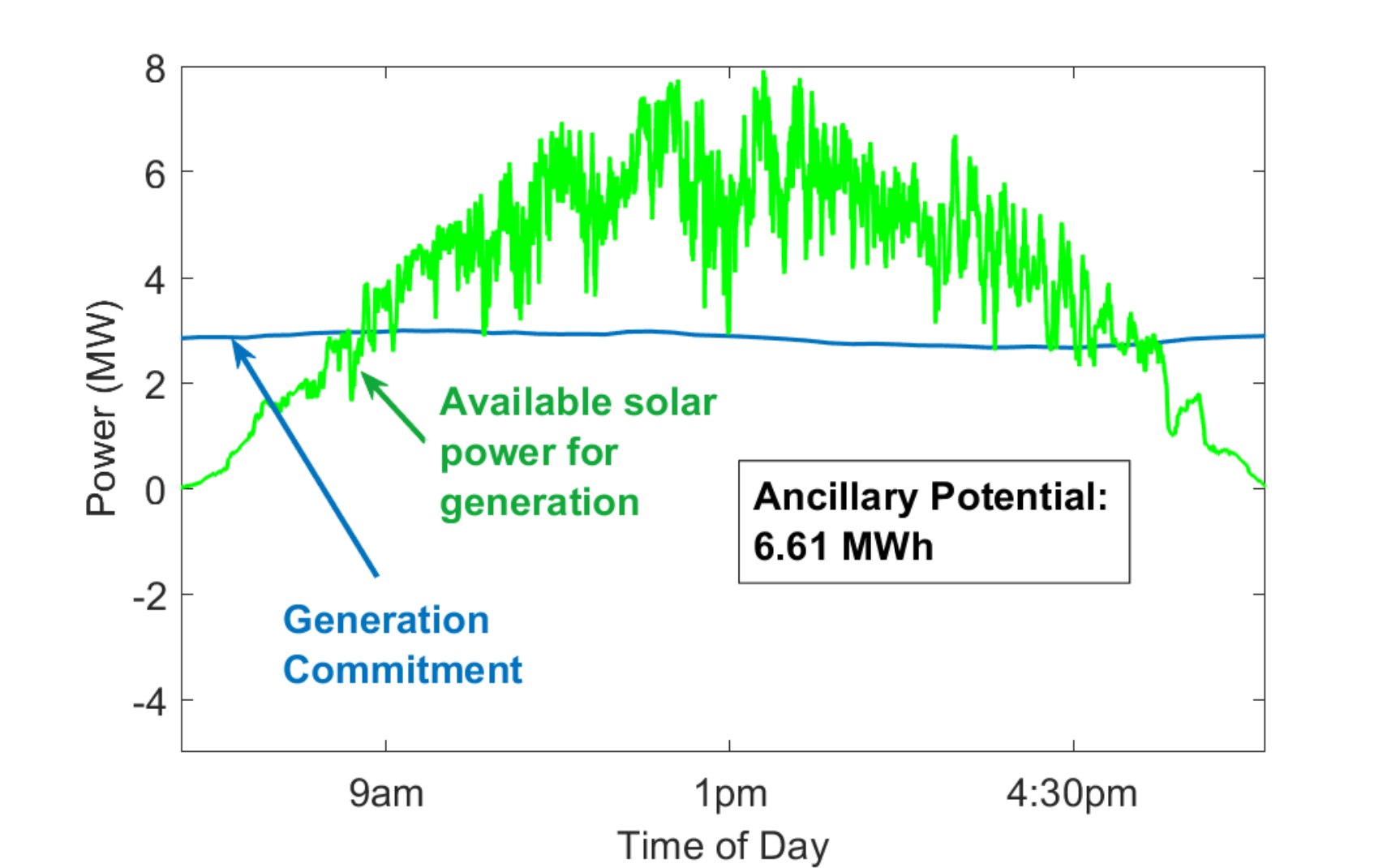}
	\caption{\label{fig:nonCurtailed} Equivalent electric power available for solar irradiance recorded on April 1, 2010, and generation commitment considered for case study.}
\end{figure}

The hierarchical control algorithm presented provides unique opportunities for solar PV plants to provide ancillary services in electricity markets. This ability consists of two separate steps: being able to proactively curtail solar generation to a fraction of the rated capacity, and remain at this committed generation level during changing weather conditions, and to be able to deviate from that set point in both directions to follow a system operator control signal. Consider the solar irradiation and generation commitment profiles presented in Fig. \ref{fig:nonCurtailed} for April 1, 2010.

The data shown in Fig. \ref{fig:nonCurtailed} indicates an ancillary potential added up to be 6.61 MWh from the day.
The ancillary potential here is computed as the area between the maximum power generation available from the solar irradiance (the MPP of the entire plant and shown by the green plot in Fig. \ref{fig:nonCurtailed}) and the generation commitment (shown by the blue plot in Fig. \ref{fig:nonCurtailed}). The main goal here is to follow the committed generation profile with as little fluctuation as possible and to precisely monitor the headroom reserve available for fulfilling instantaneous ancillary signal requests, thus demonstrating the first requirement for providing ancillary services from solar PV plants.

\begin{figure}[h]
	\centering
	\includegraphics[width=.8\linewidth]{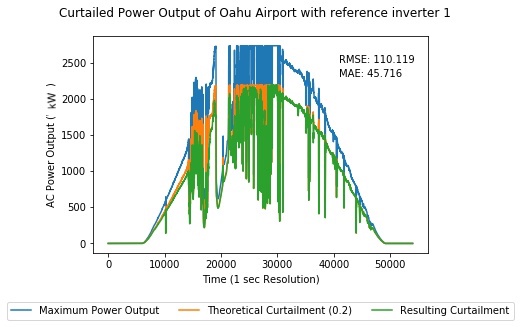}
	\caption{Simulation results of maintaining 20\% headroom for a partially cloudy day with grouping method.}
	\label{fig:Inv1Ref}
\end{figure}

As a basis of comparison with the current state-of-the-art, the grouping algorithm described in Section II is implemented to examine its performance for simply maintaining a constant 20\% headroom. These results are shown in Fig. \ref{fig:Inv1Ref} where the blue line represents the amount of power that would be produced by a solar plant operating without curtailment (uncontrolled). The yellow line is the theoretical curtailment that grouping control predicts at which it would operate. The green line is the actual power production from the solar plant after grouping control. Root mean square error (RMSE) and mean absolute error (MAE) are calculated to quantify the discrepancies between theoretical curtailment and the actual resulting curtailment. The results indicate the RMSE would be 110 kW and the MAE 46 kW. This error indicates a vast amount of untapped solar generation potential that are curtailed which is the drawback of the grouping algorithm and what the hierarchical control proposed here aims to leverage for improvement of system efficiency.

The remainder of this section presents the results in the following order. First, the decision-making process and algorithms in the proposed controller is discussed and how each layer of hierarchy contributes to achieving the overall goal of the system. Second, performance of the proposed controller for mitigating the power output fluctuations when subjected to rapid-moving clouds is compared with the grouping controller. Third, the capability of the proposed controller to respond to the dynamic frequency regulation signal (RegD signal) for providing fast frequency support is compared with that of the grouping controller.

\subsection{Inverter Signal Communication}

The proposed control system possesses a hierarchical structure and its performance relies on interaction among the control agents within layers and in between layers. This section demonstrates the performance of the virtual signaling and communication among the agents. 

\begin{figure*}[h]
	\includegraphics[width=0.9\linewidth]{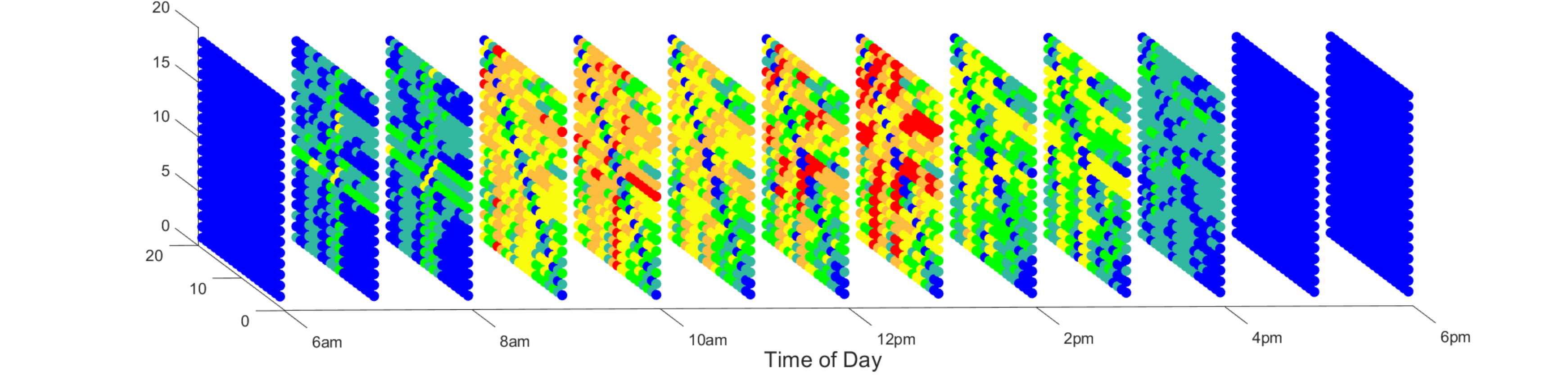}  \includegraphics[width=0.06\linewidth]{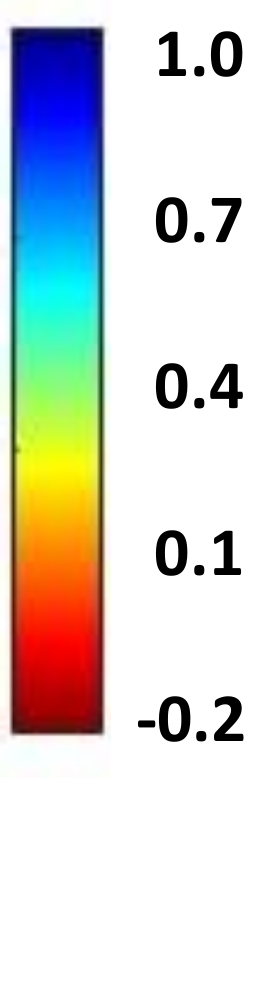}
	\centering
	\caption{\label{fig:CorrMat} Correlation heat map for all 17 inverters on April 1, 2010 for every hour of operation throughout the day. Blue signifies high direct correlation, and red suggests inverse correlation. The blue diagonal is expected because each inverter should have a correlation of 1 with itself.}
\end{figure*}

Fig. \ref{fig:CorrMat} visualizes the temporal evolution of correlation matrix for all 17 inverters considered in this case study for 8am to 6pm on April 1 in 2010. This evolution heat map shows how correlations among the inverters vary throughout the day as clouds move over the site. In this plot, red indicates the least correlated (or inversely correlated) inverters, blue indicates the most correlated inverters, and the other colors in between should be accordingly interpreted as instructed in the heatmap color guide. It should be noted that throughout the day, the diagonal elements remain consistently equal to 1 which shows the self-correlation of inverters. The varying correlation among the inverter is the backbone of decision-making in the proposed control strategy in this paper which allows to leverage the temporal and spatial variation of availability of solar across the power plant to mitigate the output fluctuations. Accordingly, the correlation matrix is updated routinely to characterize the movement of cloud coverage.

\begin{figure*}[h]
	\centering
	\begin{subfigure}[b]{0.45\textwidth}
		\centering
		\includegraphics[width=\textwidth]{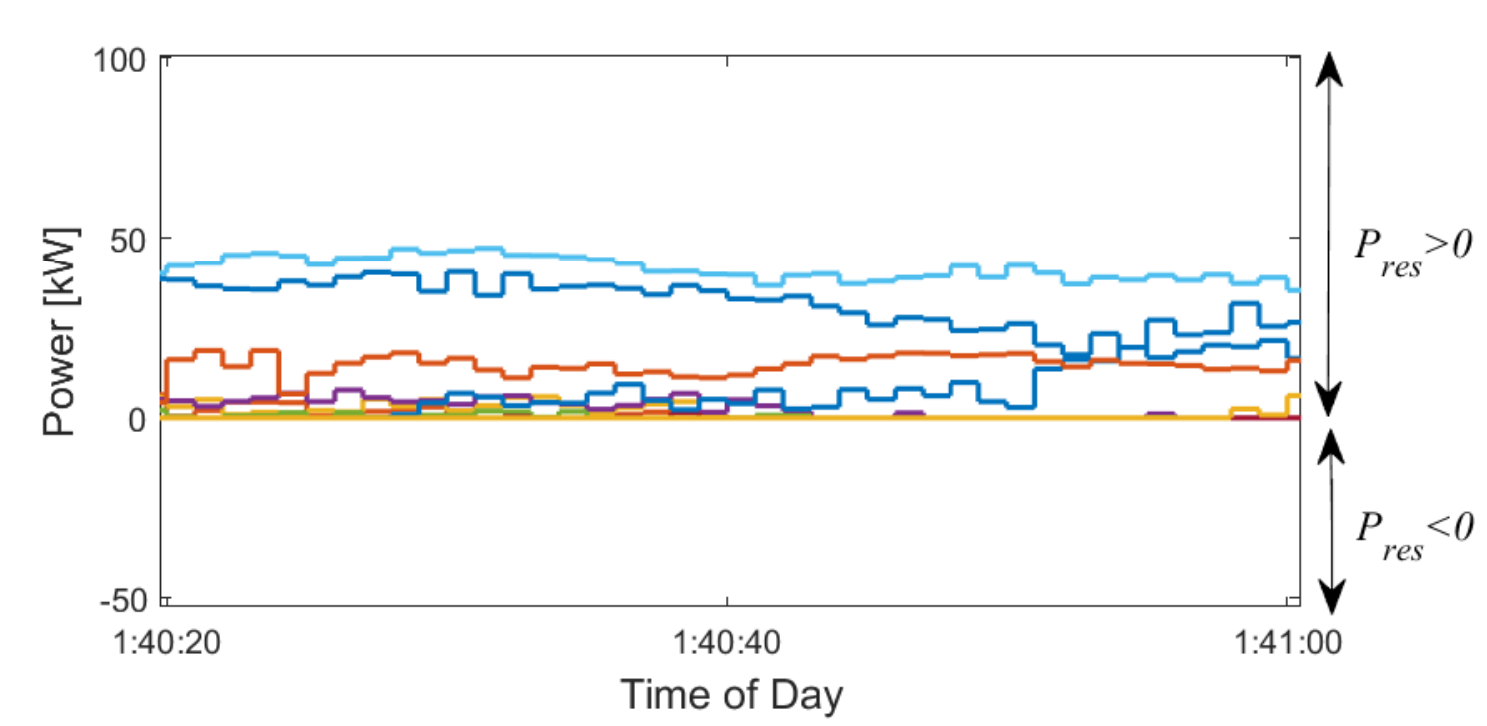}
		\caption{Not Cloudy}
		\label{noncloudy}
	\end{subfigure}
	\begin{subfigure}[b]{0.45\textwidth}
		\centering
		\includegraphics[width=\textwidth]{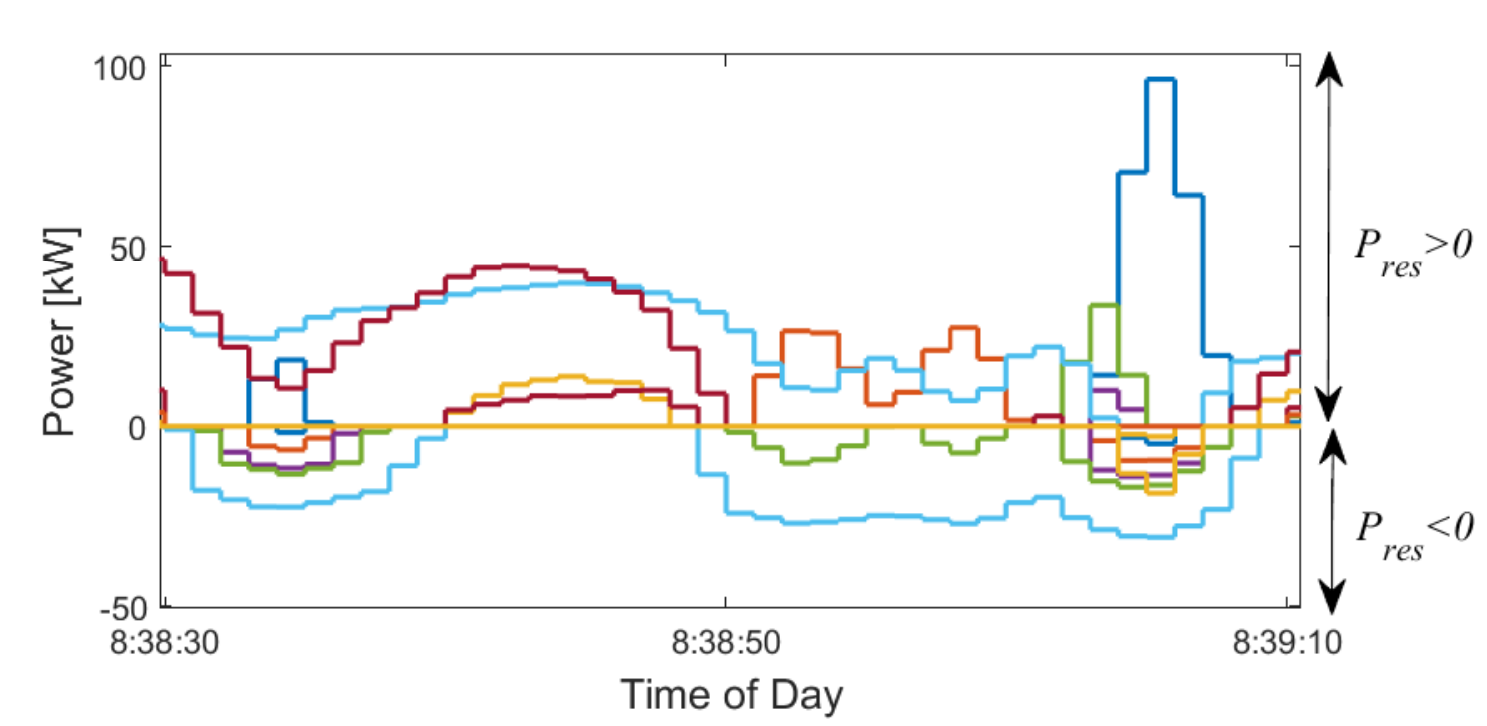}
		\caption{Cloudy}
		\label{cloudy}
	\end{subfigure}
	\caption{\label{fig:InverterCloudy} Operational power availability relative to the necessary average power output per-inverter for an entire group of inverters; the time of day in these plots is indicated by hour:minute:second.}
\end{figure*}

Fig. \ref{fig:InverterCloudy} brings further insight into the correlation among the inverters and the availability of positive residual power for individual inverters across this power plant during the cloudy and non-cloudy conditions. The not cloudy time interval in Fig. \ref{fig:InverterCloudy}a, where sunlight is available to all inverters, shows how all the inverters have positive residual power generation availability after meeting their generation obligation. Alternatively, Fig. \ref{fig:InverterCloudy}b demonstrates scarcity of residual generation reserve where some arrays are experiencing negative residuals whilst some others have positive residual power available. The negative residual values will signify to higher control levels that the inverter is in need of help from other inverters. The strategy of requesting and providing help among the control agents (PV inverters) is achieved through the signaling process as explained next.

\begin{figure*}[h]
	\centering
	\subfloat[\textbf{Output Signal:} Inverter IMPP  ]{\label{outputSig}\includegraphics[width=1\linewidth]{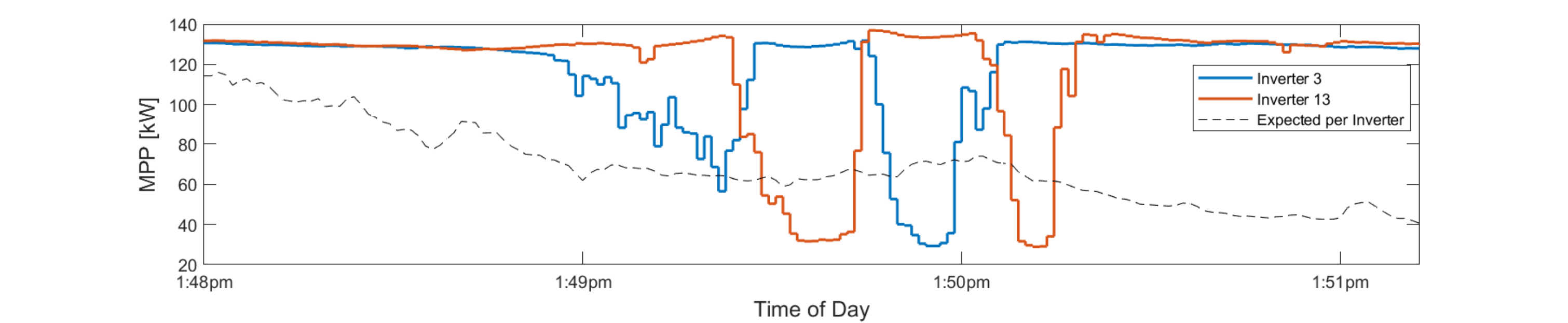}}\hfill
	\subfloat[\textbf{Return Signal:} Ratio of output to IMPP]{\label{returnSig}\includegraphics[width=1\linewidth]{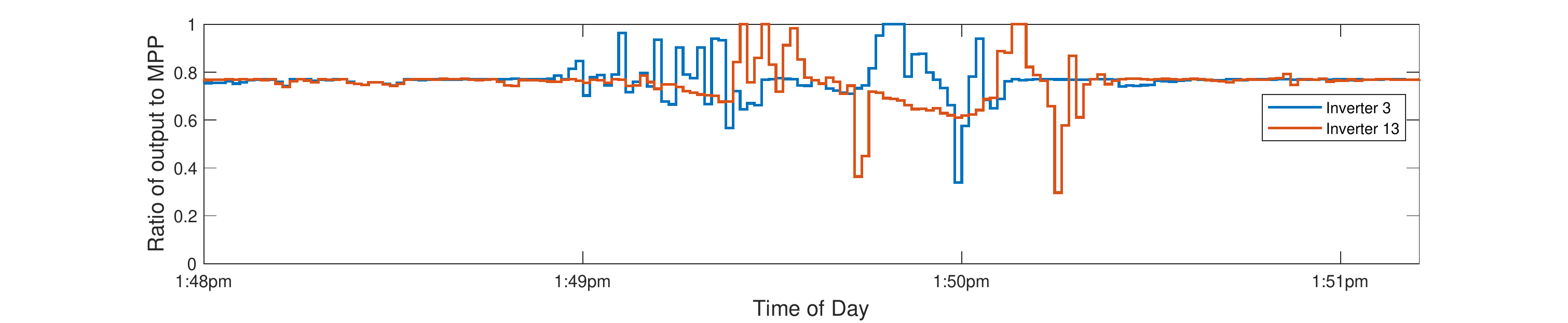}
	}
	\caption{\label{fig:InverterCom} Communication signaling results focused on only two virtually neighboring inverters over a 3 minute window of daily operation. Fig. \ref{outputSig} shows the IMPP output signal from the direct control layer. Fig. \ref{returnSig} shows the final return signal to the direct control layer of necessary curtailment.}
\end{figure*}

For demonstration, Fig. \ref{fig:InverterCom} exhibits the output signaling and return signaling of two inverter controllers (on the direct control layer) that have been chosen as virtual neighbors because of their asimilar IMPP fluctuations in the hourly statistical analysis from Figure \ref{fig:CorrMat}. The interactions between the other inverters are similar. On both ends of the time interval depicted in Fig. \ref{fig:InverterCom}a, prior to 1:49:00 and after 1:50:30 a reasonably flat maximum power point (IMPP) potential from both inverters can be seen, which suggests little to no cloud coverage over their corresponding PV panels. However, throughout the middle of the time interval, between 1:49:00 and 1:50:30, a cloud-coverage event emerges where the inverters' IMPP potentials are being reduced due to cloud cover. The IMPP status is the signal output from the direct control agent (in the direct control layer) to the supervisory controller in the layer above it and it is approximated using the expression presented in Eq. \eqref{eq:InvIMPP}. As the shading passes over the power plant over about a minute and a half, the need for help is triggered by the IMPP levels falling below the expected power generation ability per inverter (black dashed line), which is defined by the generation commitment split evenly amongst all 17 inverters.

Fig. \ref{fig:InverterCom}b shows the returned signal -- from the supervisory controller to the direct control agents -- that communicates the final information about the curtailment level at which the controller should be operating. At the direct control layer (inverter level), this can be easily translated into adjustment of the output. Although other inverters and groups within this hierarchical controller are influencing the return signal to the two inverters shown in Fig. \ref{fig:InverterCom}b as well, the results presented here make it evident that the decision-making process for these controllers are working together in a efficient and effective manner.


\begin{table}[h]
	\centering
	\footnotesize{
		\caption{Comparative results for the output regulation of the solar PV power plant for a partially cloudy day.}
		\label{tabLoadTable}
		\begin{tabular}{l|c|c|c}
			\hline
			\multicolumn{1}{c|}{\begin{tabular}[c]{@{}c@{}}\textbf{Control}\\ \textbf{Scheme} \end{tabular}} & \multicolumn{1}{c|}{\textbf{Mileage}} & \multicolumn{1}{c|}{\begin{tabular}[c]{@{}c@{}}\textbf{Unsatisfied}\\ \textbf{Commitment}  \end{tabular}}  & \multicolumn{1}{c}{\begin{tabular}[c]{@{}c@{}}\textbf{Satisfied}\\ \textbf{Commitment}  \end{tabular}} \\
			\hline
			Grouping             &  245 kW    & 8,089 kWh  & 88.36\%                                   \\ \hline
			Hierarchical &   14 kW  &435 kWh   & 99.35\%                                                      \\ \hline
		\end{tabular}
}
\end{table}

\subsection{Satisfying the Generation Commitment}

\begin{figure*}[h]
	\centering
	\subfloat[Grouping control (the algorithm proposed in \cite{vahan})]
	{\label{groupingR}\includegraphics[width=1\linewidth]{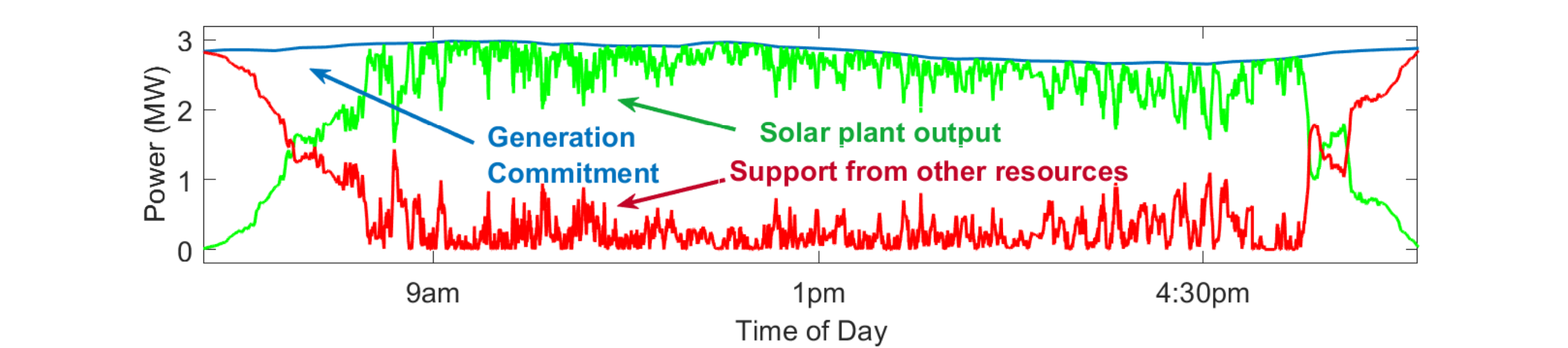}}
	\hfill
	\subfloat[Proposed hierarchical control]
	{\label{hierarchicalR}\includegraphics[width=1\linewidth]{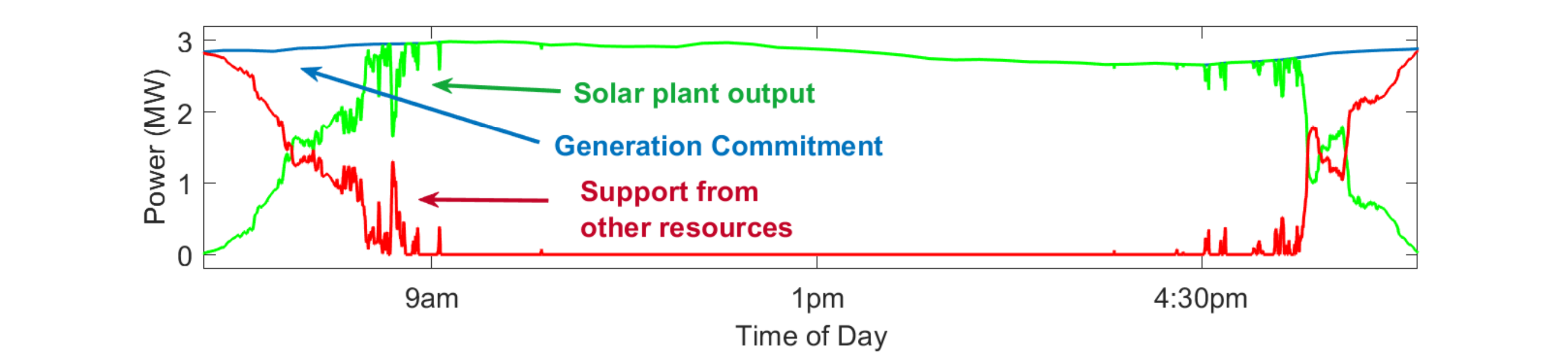}
	}
	\caption{\label{fig:bigResult} Comparison of performance by grouping control vs. proposed hierarchical control in delivering generation commitment by the solar power plant considered in this case study. The red line represents supportive power generation that is required to meet generation commitment.}
\end{figure*}

Fig. \ref{fig:bigResult} presents the results for the capability of the proposed hierarchical controller for the mitigation of power fluctuations and following the power generation commitment. The results are compared with that of the grouping controller in the same plot. Mitigating the fluctuations shown by the red line is what will allow utilities to operate solar power plants without the need for additional support from additional resources (storage or fast responding hybrid resources) or stressing the network.  The results here are visual evidence of the superiority of the proposed controller over the grouping controller by almost ideally following the generation obligation curve.

To quantitatively assess the performance of the both control algorithms considered, Table \ref{tabLoadTable} presents the results using three common metrics: (1) Mileage in kW, (2) Regulation in kWh, and (3) Commitment Satisfied as a percentage. These metrics are important measures in evaluation of generation fleet in an energy market. Mileage is a metric used to measure the instantaneous ramping that the solar power plant would require of supportive generation and/or energy storage \cite{8974891}. Reducing mileage on the system has value because it will extend the lifetime of a battery and minimize fuel costs of fossil fuel generators. Regulation is the total amount of energy that was supplied from generation sources other than the solar power plant in order to maintain the load demand \cite{choi2016hybrid}. Reducing the amount of regulation also minimizes fuel costs (and thus carbon emissions). Lower mileage and regulation means significantly less operational costs. Commitment satisfaction is the percentage success rate of satisfying the precise generation amount that the power plant has committed to. Having a low satisfied commitment percentage would risk fines for a generator providing ancillary services, and eventually remove the generator from consideration in an ancillary service market.
The result presented in Table \ref{tabLoadTable} yielded far less mileage and regulation throughout the day for the proposed hierarchical control system in comparison with the grouping control, as well as approximately 18 times less unsatisfied commitment. Hierarchical control also satisfied nearly 100\% of its generation commitments with precision, which was 11\% more than the grouping control method. In practice, this will result in reduced cost of operation and increased market efficiency and reliability.

The results presented in Fig. \ref{fig:bigResult} and Table \ref{tabLoadTable} clearly indicate the effectiveness of the proposed hierarchical control in satisfying generation commitment. Having established that the proposed approach can effectively curtail during changing weather conditions, its ability to respond to an ancillary service signal request is simulated in the following subsection.

\subsection{Response to Ancillary Service Request}
The ability for a utility-scale solar power plant to participate in the ancillary service market is a potential additional source of revenue for PV plants that is likely to become more common with increasing penetration levels of solar power. Here, the solar power plant of the case study was subjected to a dynamic frequency response signal for the hours between 9am to 5pm in between which the ancillary service potential was available in both the upward and downward directions. The hours considered are the period that the solar availability was greater than the generation commitment, when the solar PV was curtailed to satisfy the generation commitment as a first priority. The dynamic ancillary service signal is Regulation D, commonly referred to as RegD, is a fast frequency ancillary request signal from the PJM market which is sent to participating units every 2 seconds. 

For this simulation, the RegD signal time-series was received directly by the adaptive layer of the control system. Subsequently, the command from the adaptive layer then flows through the hierarchy of decision making in order to set the plant at the desired operational level. The performance of the proposed hierarchical controller and the grouping controller for an identical day of RegD signals are evaluated and shown in Table \ref{tabRegDTable}. In this table, the metric used to measure the plant response to the RegD signal is the amount of RegD signals that are fulfilled entirely. Typically, fossil generators are required to satisfy more than 75\% of requests in PJM.

\begin{table}[h]
	\centering
	\caption{Comparative results for response of the power plant output to the dynamics frequency response signal}
	\label{tabRegDTable}
	\begin{tabular}{l|c}
		\hline
		
		\textbf{Control Scheme }               & \textbf{Satisfied RegD Signal}\\ \hline
		Grouping Control           &81.45\% \\ \hline  
		Hierarchical Control & 99.88\%\\ \hline  
	\end{tabular}
\end{table}

The results presented in Table \ref{tabRegDTable} shows superior performance of the proposed hierarchical controller when compared to the grouping controller. The success-rate of hierarchical controller is more than 18\% greater than the grouping controller in responding to the requests sent by RegD signal. 


\section{Conclusion}

This paper proposed a hierarchical control structure and algorithm for curtailment of solar PV power plants. The control goals were to (1) mitigate the variability of the plant power output caused by the movement of clouds over the plant, and (2) maintain a desired headroom reserve to provide the plant with the capability to participate in ancillary service provisions of the grid. The proposed control system was implemented on a case study from a site in Hawaii and was compared with the current technology which is grouping control algorithm. The results indicated that the proposed control strategy was able to satisfy 99.35\% of generation commitment and 99.88\% of ancillary service signal requests. Relative to the current state of the art, a grouping control method, the proposed hierarchical control algorithm operated with 18 times less mileage and unsatisfied generation, which could largely expand the lifetime of any connected storage system by optimizing charging cycles.  

This proposed control system can set a basis for enabling technologies to achieve a carbon-free electric power sector. Next stages of this work could include a more detailed execution of this algorithm by embedding the software into a physical test solar power plant. Additionally, the formulation of the algorithm could be expanded upon to interface intelligently with a battery storage system to optimize storage for the health of the battery, stability needs of the power grid, and profitability in ancillary and real-time energy markets. Lastly, with potential application to controlling wind power plants as well, this hierarchical control structure could be a necessary piece of a larger solution to developing a highly automated grid operating system that controls manages multiple power plants at once.



\section{Acknowledgment}
The authors wish to thank Nick Parker from Southwest Power Pool (SPP) for sharing the recent generation information. 

This work was financially supported by the U.S. Department of Energy (DOE) under Contract No. DE-AC36-08GO28308.


\begin{thebibliography}{10}
	
	\bibitem{8528319}
	O.~{Ogunrinde}, ``Investing in renewable energy: Reconciling regional policy
	with renewable energy growth,'' {\em IEEE Engineering Management Review},
	vol.~46, no.~4, pp.~103--111, 2018.
	
	\bibitem{Lazard2020}
	{Lazard Consultant}, {\em {Levelized Cost of Energy and Levelized Cost of
			Storage 2020}}, 2020 (accessed February 8, 2020).
	
	\bibitem{Biden}
	{The White House: Briefing Room}, {\em {2030 Greenhouse Gas Pollution Reduction
			Target Aimed at Creating Good-Paying Union Jobs and Securing U.S. Leadership
			on Clean Energy Technologies}}, April 22, 2021.
	
	\bibitem{Countries}
	``Which countries have legally-binding net-zero emissions targets?,'' {\em NS
		Energy}.
	
	\bibitem{olek2014deployment}
	B.~Olek, ``Deployment of energy storages for ancillary services,'' in {\em 11th
		International Conference on the European Energy Market (EEM14)}, pp.~1--5,
	IEEE, 2014.
	
	\bibitem{lee2008small}
	D.-J. Lee and L.~Wang, ``Small-signal stability analysis of an autonomous
	hybrid renewable energy power generation/energy storage system part i:
	Time-domain simulations,'' {\em IEEE Transactions on energy conversion},
	vol.~23, no.~1, pp.~311--320, 2008.
	
	\bibitem{bravo2018integration}
	R.~Bravo and D.~Friedrich, ``Integration of energy storage with hybrid solar
	power plants,'' {\em Energy Procedia}, vol.~151, pp.~182--186, 2018.
	
	\bibitem{paska2009hybrid}
	J.~Paska, P.~Biczel, and M.~K{\l}os, ``Hybrid power systems--an effective way
	of utilising primary energy sources,'' {\em Renewable energy}, vol.~34,
	no.~11, pp.~2414--2421, 2009.
	
	\bibitem{hu2006hybrid}
	W.~Hu, Q.~Ruan, W.~Wang, S.~Mei, and Q.~Lu, ``Hybrid power control system and
	its application,'' in {\em 2006 International Conference on Power System
		Technology}, pp.~1--5, IEEE, 2006.
	
	\bibitem{sajadi2019power}
	A.~Sajadi, K.~Loparo, {\L}.~Roslaniec, and M.~K{\l}os, ``Power sharing based
	control of hybrid wind-diesel standalone systems,'' in {\em IEEE EUROCON
		2019-18th International Conference on Smart Technologies}, pp.~1--5, IEEE,
	2019.
	
	\bibitem{NELSON2020114963}
	J.~R. Nelson and N.~G. Johnson, ``Model predictive control of microgrids for
	real-time ancillary service market participation,'' {\em Applied Energy},
	vol.~269, p.~114963, 2020.
	
	\bibitem{bohnet2017hybrid}
	B.~Bohnet, S.~Kochanneck, I.~Mauser, S.~Hubschneider, M.~Braun, H.~Schmeck, and
	T.~Leibfried, ``Hybrid energy storage system control for the provision of
	ancillary services,'' in {\em International ETG Congress 2017}, pp.~1--6,
	VDE, 2017.
	
	\bibitem{clastres2010ancillary}
	C.~Clastres, T.~H. Pham, F.~Wurtz, and S.~Bacha, ``Ancillary services and
	optimal household energy management with photovoltaic production,'' {\em
		Energy}, vol.~35, no.~1, pp.~55--64, 2010.
	
	\bibitem{clastres2010optimal}
	C.~Clastres, T.~H. Pham, F.~Wurtz, and S.~Bacha, ``Optimal household energy
	management and participation in ancillary services with pv production,'' {\em
		Energy}, vol.~35, no.~1, pp.~55--64, 2010.
	
	\bibitem{berrada2016operation}
	A.~Berrada and K.~Loudiyi, ``Operation, sizing, and economic evaluation of
	storage for solar and wind power plants,'' {\em Renewable and sustainable
		energy Reviews}, vol.~59, pp.~1117--1129, 2016.
	
	\bibitem{saez2016co}
	A.~Saez-de Ibarra, A.~Milo, H.~Gaztanaga, V.~Debusschere, and S.~Bacha,
	``Co-optimization of storage system sizing and control strategy for
	intelligent photovoltaic power plants market integration,'' {\em IEEE
		Transactions on Sustainable Energy}, vol.~7, no.~4, pp.~1749--1761, 2016.
	
	\bibitem{saez2016sizing}
	A.~Saez-de Ibarra, E.~Martinez-Laserna, D.-I. Stroe, M.~Swierczynski, and
	P.~Rodriguez, ``Sizing study of second life li-ion batteries for enhancing
	renewable energy grid integration,'' {\em IEEE Transactions on Industry
		Applications}, vol.~52, no.~6, pp.~4999--5008, 2016.
	
	\bibitem{saez2016management}
	A.~Saez-de Ibarra, V.~I. Herrera, A.~Milo, H.~Gaztanaga, I.~Etxeberria-Otadui,
	S.~Bacha, and A.~Padros, ``Management strategy for market participation of
	photovoltaic power plants including storage systems,'' {\em IEEE Transactions
		on Industry Applications}, vol.~52, no.~5, pp.~4292--4303, 2016.
	
	\bibitem{lee2018optimal}
	J.-Y. Lee, K.~B. Aviso, and R.~R. Tan, ``Optimal sizing and design of hybrid
	power systems,'' {\em ACS Sustainable Chemistry \& Engineering}, vol.~6,
	no.~2, pp.~2482--2490, 2018.
	
	\bibitem{lee2020sizing}
	J.~Y. Lee, Y.~C. Lu, and Y.~C. Chen, ``A sizing-validation approach to hybrid
	power system design and planning,'' {\em Process Safety and Environmental
		Protection}, vol.~141, pp.~178--189, 2020.
	
	\bibitem{conte2018mixed}
	F.~Conte, F.~D'Agostino, P.~Pongiglione, M.~Saviozzi, and F.~Silvestro,
	``Mixed-integer algorithm for optimal dispatch of integrated pv-storage
	systems,'' {\em IEEE Transactions on Industry Applications}, vol.~55, no.~1,
	pp.~238--247, 2018.
	
	\bibitem{8370779}
	X.~{Chen}, ``Forecasting-based power ramp-rate control strategies for
	utility-scale pv systems,'' {\em IEEE Trans. on Industrial Electronics},
	vol.~66, no.~3, pp.~1862--1871, 2019.
	
	\bibitem{vahan}
	V.~Gevorgian, ``Highly accurate method for real-time active power reserve
	estimation for utility-scale photovoltaic power plants,''
	
	\bibitem{vahan2}
	V.~Gevorgian and B.~O'Neill, ``Advanced grid-friendly controls demonstration
	project for utility-scale pv power plants,''
	
	\bibitem{loutan2017demonstration}
	C.~Loutan, P.~Klauer, S.~Chowdhury, S.~Hall, M.~Morjaria, V.~Chadliev,
	N.~Milam, C.~Milan, and V.~Gevorgian, ``Demonstration of essential
	reliability services by a 300-mw solar photovoltaic power plant,'' tech.
	rep., National Renewable Energy Lab.(NREL), Golden, CO (United States), 2017.
	
	\bibitem{byrne2016estimating}
	R.~H. Byrne, R.~J. Concepcion, and C.~A. Silva-Monroy, ``Estimating potential
	revenue from electrical energy storage in pjm,'' in {\em 2016 IEEE Power and
		Energy Society General Meeting (PESGM)}, pp.~1--5, IEEE, 2016.
	
	\bibitem{PJM}
	{Day-Ahead and Real-Time Market Operations}, {\em {PJM Manual 11: Energy \&
			Ancillary Services Market Operations Revision: 113}}, 2021.
	
	\bibitem{rosu2013practical}
	M.~Rosu-Hamzescu and S.~Oprea, ``Practical guide to implementing solar panel
	mppt algorithms,'' {\em Microchip Technology Inc, Application Note, AN1521},
	2013.
	
	\bibitem{data}
	M.~Sengupta and A.~Andreas, ``Oahu solar measurement grid (1-year archive):
	1-second solar irradiance; oahu, hawaii (data),'' {\em NREL Report No.
		DA-5500-56506.}, 2010.
	
	\bibitem{PVLib2}
	{Holmgren, William and Hansen, Clifford and Mikofski, and Mark}, {\em {pvlib
			python: a python package for modeling solar energy systems.}}, 2019.
	
	\bibitem{8974891}
	X.~Ma, R.~Xu, C.~Wei, W.~Kang, and M.~Chen, ``Power mileage-based allocation
	and dispatch strategy of battery energy storage system,'' in {\em 2019 IEEE
		Sustainable Power and Energy Conference (iSPEC)}, pp.~1486--1491, 2019.
	
	\bibitem{choi2016hybrid}
	J.~W. Choi, S.~Y. Heo, and M.~K. Kim, ``Hybrid operation strategy of wind
	energy storage system for power grid frequency regulation,'' {\em IET
		Generation, Transmission \& Distribution}, vol.~10, no.~3, pp.~736--749,
	2016.
	
\end{thebibliography}

\end{document}